\documentclass[iop,apj]{emulateapj}

\slugcomment{Draft Version July 31, 2013}

\shorttitle{The Simulated Cloud Catalogue}
\shortauthors{Benincasa et al.}

\begin{document}

\title{Giant Molecular Cloud Formation in Disk Galaxies:  Characterizing Simulated versus Observed Cloud Catalogues}

\author{Samantha M. Benincasa\altaffilmark{1}, Elizabeth J. Tasker\altaffilmark{2,3}, Ralph E. Pudritz\altaffilmark{1}, and James Wadsley\altaffilmark{1}}

\altaffiltext{1}{Department of Physics and Astronomy, McMaster University, Hamilton, ON, L8S 4M1, Canada}
\altaffiltext{2}{Department of Physics, Faculty of Science, Hokkaido University, Kita-ku, Sapporo 060-0810, Japan}
\altaffiltext{3}{CITA National Fellow - McMaster University}

\begin{abstract}

We present the results of a study of simulated Giant Molecular Clouds (GMCs) formed in a Milky Way-type galactic disk with a flat rotation curve.  This simulation, which does not include star formation or feedback, produces clouds with masses ranging between $10^4 {\rm M}_{\odot}$ and $10^7 {\rm M}_{\odot}$.  We compare our simulated cloud population to two observational surveys; The Boston University- Five College Radio Astronomy Observatory Galactic Ring Survey and the BIMA All-Disk Survey of M33.  An analysis of the global cloud properties as well as a comparison of Larson's scaling relations is carried out.  We find that simulated cloud properties agree well with the observed cloud properties, with the closest agreement occurring between the clouds at comparable resolution in M33.  Our clouds are highly filamentary - a property that derives both from their formation due to gravitational instability in the sheared galactic environment, as well as to cloud- cloud gravitational encounters.  We also find that the rate at which potentially star forming gas accumulates within dense
regions - wherein $n_{\rm thresh} \ge 10^4$\,cm$^{-3}$ -  is  3$\%$ per 10 Myr, in clouds of roughly $10^6 M_{\odot} $.  This suggests that star formation rates in observed clouds are related to the rates at which gas can be accumulated into dense subregions within GMCs via filamentary flows.   The most internally well-resolved clouds are chosen for listing in a catalogue of simulated GMCs; the first of its kind. The catalogued clouds are available as an extracted data set from the global simulation.
\end{abstract}

\keywords{Galaxies: ISM; Galaxies: spiral; ISM: clouds; ISM: structure; Methods: numerical }


\section{Introduction}

Stars form  in the secluded depths of the densest cores in cold interstellar clouds. These extended structures of molecular hydrogen and dust are galactic wombs for their stellar populations and it is their environments that control the creation rate and properties of the new stars. Therefore, in order to explore any question regarding star formation, we must  ultimately account for the properties of these stellar cradles that will be their birth place. 

The molecular hydrogen clouds are collectively referred to as the Giant Molecular Clouds (GMCs). The average gas density of these structures is a few hundred atoms per ${\rm cm}^3$, with stars forming within a star cluster in cores that exceed densities of ${\rm n_{H}} > 10^4 \ {\rm cm}^{-3}$ \citep{Lada2010}. While the cores themselves are clearly undergoing gravitational collapse, the gravitational binding of the cloud as a whole is more debatable, with both observations and simulations disagreeing as to which side of the virial line a cloud is likely to sit \citep{hopkins, Dobbs_virial, Tasker2011, Tasker2009, Hirota2011, Heyer2009,  McKee2007, Rosolowsky2007}.

One of the main reasons why the binding of the cloud is not clear is that the cloud's formation is not a straight forward process. The galactic environment in which a GMC forms is rife with instabilities and shocks that move through a multiphase, turbulent medium that shapes, stirs and buffets the cloud. This means that attempts to explore the star forming environment assuming a simple distribution of gas may fall far short of the true internal properties of a GMC formed in the above conditions. This has implications for numerical studies on the scale of star forming cores.  To what extent then, do the local properties of star formation within highly bound subregions such as dense cores and clumps follow from the larger scale internal dynamics of the clouds in which such entities are formed?   The answer to this question is particularly important for numerical studies since the highest resolutions available to chart the formation and evolution of clouds and their substructure are barely adequate to spatially resolve the largest, cluster forming clumps - let alone the details of individual star forming cores. 

Despite the complexity of their formation process, many properties of the GMCs are well described by observations, both in the empirical relations relating mass, size and velocity dispersion ~\citep{larson1981} and in global observational studies of GMC populations in the Milky Way \citep{Heyer2009, duval}, and nearby galaxies such as M33 \citep{M33}, IC342 \citep{Hirota2011}, M64 \citep{Rosolowsky2005}, M31 \citep{Rosolowsky2007} and the LMC \citep{Kawamura2009}. These provide handles to quantitatively assess whether a GMC created in simulations is a realistic star formation environment. If the object passes these tests, then it can be used confidently to study the initial conditions for the star formation process. One of the major goals of this paper is to create a catalogue of GMCs that agree with the observational data that can be used as a comparison tool for observers and theorists alike and can potentially form the starting point for further higher resolution studies in star formation. For this, we will focus on two observational surveys in particular; the Boston University- Five College Radio Astronomy Observatory Galactic Ring Survey \citep{Heyer2009} and the BIMA All-Disk Survey of M33 \citep{M33}.  Another important application of our simulations is to measure the rate at which dense, star forming gas develops within our most massive (i.e., best spatially resolved ) GMCs without the complicating effects of stellar feedback.

\subsection{Numerical Studies of GMCs}
There are two main theories concerning the initial formation of GMCs; the `top-down' model wherein clouds are formed as a result of global instabilities in the sheared, self-gravitating interstellar medium of the galactic disk and the `bottom-up' scenario where clouds are formed by agglomeration or coagulation \citep[][and references therein]{McKee2007}. The actual physical mechanism may involve a number of processes, but due to time-scale restraints, the former `top-down' method has been more widely investigated.  It is this approach that will be addressed in this paper. 

The `top-down' category can be further split by asking what causes the gathering of gas into a GMC. For this, there are also two main candidates; buoyancy and self-gravity. The former of these centres around the interplay between gravity (which wants to compress the gas in the galactic mid-plane) and the magnetic field (which acts to gather gas at the galactic footpoints of rising loops of magnetic field) in the Parker instability. 
Specifically the Parker instability produces the lower mass clouds of a
few 10$^5$  M$_{\odot}$ (where the gas is not very self gravitating), while self gravity produces the higher mass clouds (a few 10$^6$ M$_{\odot}$), where magnetic fields have little effect\citep{elmegreen1982}. The added compression need not necessarily be due to a spiral wave, since the sheared galactic medium can also act with weak magnetic fields to produce magneto-rotational instabilities (MRI) in the galactic disk \citep{sellwood1999}. Such MRI instabilities have also been shown to seed GMC growth in local-scale simulations \citep{kim}. Once initiated, the Toomre instability of the self-gravitating sheared gas was found to rapidly build massive GMCs, reaching $\gtrsim 10^7\,{\rm M}_\odot$ .

The self-gravity, top-down formation has been explored on both global and local scales. On the local scale, colliding flows have been used to create the clouds. These are counted in the `top-down' camp since the flows themselves can be driven by gravitational instabilities such as spiral shocks or cloud-cloud collisions, although may also have other sources such as supernovae or stellar winds. \cite{Heitsch2008} and \cite{banerjee} have performed studies of this kind, where clouds form at the interface of two colliding flows of gas, naturally accounting for the turbulent nature of the GMCs. \cite{superbubble} utilised the idea of colliding wind-blown superbubbles, finding both cold gas clumps and filaments formed at the interface. 

On the global scale, the entire galactic disk is included in these simulations.  The disk becomes gravitationally unstable via the Toomre instability and fragments to form the GMCs \citep{Tasker2011, Dobbs_virial, Bournaud2010, Tasker2009, Agertz2009, Dobbs_agg}. This scenario has the advantage of naturally including the effects of shear and also interactions between neighbouring clouds, which drives the cloud's internal turbulence \citep{Dobbs_virial} and may play an important role in the star formation process \citep{Tasker2011, Tasker2009, Tan2000}. Once formed, the clouds grow by accretion and cloud-cloud collisions.  Star formation is among the mechanisms that would destroy or shred such clouds.  

In the absence of star formation, \citet{Tasker2009} found good agreement between the properties of the created GMC population and those from observations, including mass, radius, velocity dispersion and rotation. They also found a high cloud collision rate, with clouds merging approximately every $1/5$th of an orbital period. Such a result suggests that cloud collisions are an important component of  achieving realistic simulated GMCs, emphasising the importance of including the global environment in their formation. 

Simulations on this scale that have followed the GMC formation process through to include the production of stars and their resultant feedback, find that these internal processes can dictate the GMC properties at later times. In the presence of a spiral potential, \citet{Dobbs_virial} found that without feedback, clouds tend to be spherical and bound while they become unbound when feedback is included. This result was repeated by \citet{hopkins} who compared the impact of different forms of feedback and found that while no feedback produced very tightly bound clumps, there was a wide variation when feedback was included, with larger clouds tending to be more bound.

Overall, the top-down formation mechanism for GMCs has proved to be highly successful in numerical simulations, both with and without the inclusion of magnetic fields and stellar feedback. However, which roles dominate during the star formation process and cloud properties such as their gravitational boundedness remain an active topic of discussion.  

The success of the Tasker \& Tan (2009) model, despite its simplicity in the absence of feedback or magnetic fields, suggests that gravity may be the controlling force in the GMCÕs early evolution. With this in mind, we adopt a similar model (although with a few notable changes laid out in Section ¤2) to explore the pre-star formation and evolution of GMCs through a top-down, self-gravitating mechanism in the absence of star formation and localised feedback. This allows us to both study the interactions and formation of GMCs in a purely dynamical sense and measure the rate at which dense, potentially star forming gas, develops within the clouds.


\begin{figure*}
\centering
\includegraphics[width=180mm,height=55mm]{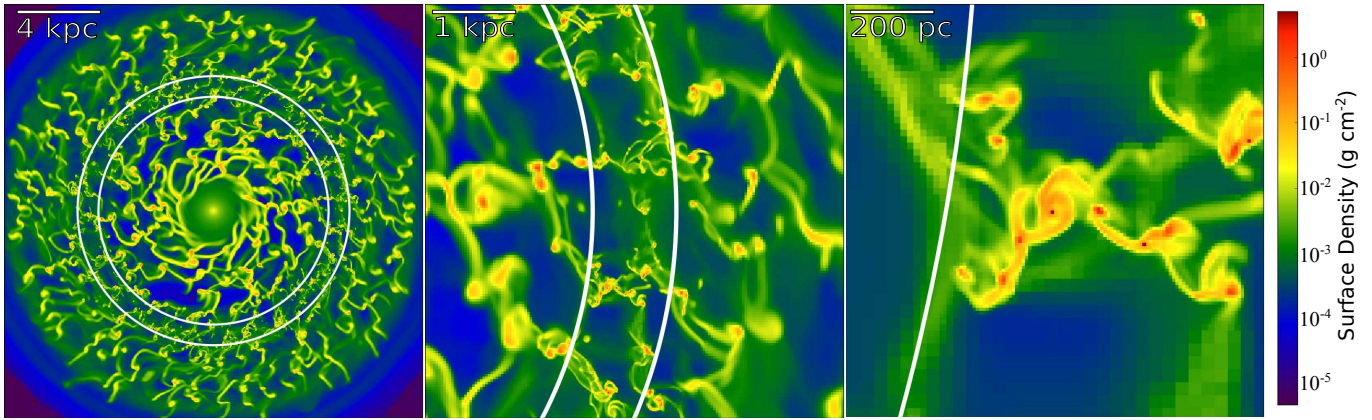}
\caption{Three surface density projections of a galactic disk at different scales.  \textbf{Left:} The full disk, spanning $20$\,kpc.  The maximally refined co-rotating region, i.e. the region from which clouds are harvested, is within the white outlined ring.  \textbf{Centre:} Zoom-in, image is approximately $5$\,kpc across.  Here again, the co-rotating region is within the white outline.  At this scale a clear difference is visible between our co-rotating region and the rest of the disk.  Clouds are the most refined and finely structured in this region.  \textbf{Right:} Close-up of a two clouds identified by the friends-of-friends algorithm; the image is $1$\,kpc across. Images were rendered using yt, an astrophysics analysis and visualization package \citep{yt}}
\label{fig:zoom_scale}
\end{figure*}

\subsection{Observational Surveys of GMCs}

We compare our simulated clouds with two major observational surveys. The first of these is the Boston University - Five College Radio Astronomy Observatory Galactic Ring Survey (BU-FCRAO)~\citep{Heyer2009,  duval} (hereafter, GRS), which uses the tracer $^{13}$CO (J = 1 $\rightarrow$ 0) . This survey re-examined the clouds found in the seminal \citet{solomon} study of the GMCs in the galactic ring. These new results have an increased resolution of $0.2$\,pc at a distance of $1$\,kpc and utilised the CLUMPFIND automated algorithm\,\citep{clumpfind} to identify the clouds within the sample. 

Our second comparison survey is the BIMA All-Disk survey, which uses the tracer  $^{13}$CO (J = 1 $\rightarrow$ 0) and looks at extragalactic clouds in M33\,\citep{M33}. The inclusion of this second survey has two major advantages: the first is that comparisons with both galactic and extragalactic clouds allows more general conclusions to be drawn about the nature of the GMCs. Secondly, this survey has a spatial resolution of $20$\,pc, which is close to our own resolution limit (see Section~\S2), making it the more comparable data set. Additional support for our top-down, gravitational fragmentation model is found in their paper, with \citet{M33} suggesting that the clouds in M33 are most likely to have been formed through instabilities, rather than agglomeration and further suggest that the properties of the GMCs are determined by the mass of the cloud. 

In this paper we present an analysis of clouds harvested from a simulation of a Milky Way-type disk.  In Section~\S2, we discuss the simulation and method of GMC analysis. Section~\S3 looks at the structure of the GMCs and in Sections~\S4 and \S5 we compare the cloud properties to two observational surveys of clouds in the Milky Way and M33. Finally in Section~\S6, we present a numerical catalogue of these GMCs and propose that these can be used as realistic initial conditions for higher resolution studies of massive star and star cluster formation. Section~\S7 details our conclusions. 


\section{Numerical Methods}

The clouds analysed in this paper are harvested from an isolated global galaxy disk simulation of a Milky Way-type galaxy. The simulation was performed using \textit{Enzo}; a three-dimensional adaptive mesh refinement hydrodynamics code \citep{enzo1, enzo2, enzo3}. The disk is contained in a three-dimensional box of side length 32 kpc with a root grid of $128^3$ and five possible levels of refinement. This results in a limiting resolution of 7.8 pc. The disk itself is kept at a minimum resolution of 31 pc, with three levels of refinement permanently placed around that volume. This ensures the disk is stable in the absence of cooling. \textit{Enzo} evolves the gas using a three-dimensional version of the \textit{Zeus} hydrodynamics algorithm \citep{zeus}. This routine uses an artificial viscosity term to represent shocks, where the variable associated with this, the quadratic artificial viscosity, was set to 2.0 (the default value). We include a radially dependent photoelectric heating term described in detail in \citet{Tasker2011}, but no other star formation or localised feedback. Radiative cooling is allowed down to $300$\,K, a value that corresponds to a minimum velocity dispersion of $1.8$\,km/s. This limit crudely allows for processes that are below our resolution limit, such as the internal turbulence of the cloud on scales below $7.8$\,pc and pressure from magnetic fields. The cooling follows the analytical curve of \citet{Sarazin1987} to $T = 10^4$\,K and the extension to $300$\,K from \citet{Rosen1995}.

In addition to this first simulation, a second simulation was completed wherein a pressure floor was imposed. This floor acted in such way that when the Jeans length became less than four times the cell size (and therefore the simulation became unresolved according to the \citet{Truelove1997} criterion), an additional pressure term was added that caused the gas to follow a polytrope with an adiabatic index of $\gamma = 2$. This brought the collapse to a halt and prevented the creation of isolated single cells with extremely high density that are not resolved by the simulation. In fact, the inclusion of this term produced identical results for the cloud bulk properties of mass, radius and surface density and only affected the velocity dispersion due to the effect of the increased pressure on the cloud's internal dynamics. As this pressure injection is a numerical addition, the resultant velocity dispersion did not hold physical meaning. Since we do ultimately expect the gas within the cloud to form a dense core during star formation, we have opted to present the first simulation without the pressure floor in this paper, but note that the internal dynamics of the cloud are poorly resolved at this resolution. 

The disk is initially borderline gravitationally stable, fragmenting as the gas cools to form dense clouds of gas that we identify as the giant molecular clouds. The initial disk mass is 6.5$\times10^9 M_{\odot}$.  Full details of the initial set-up of the simulation are given in \cite{Tasker2009, Tasker2011}. 

We developed a special procedure to create clouds with high spatial resolution.  In difference to the previous simulations of \cite{Tasker2009, Tasker2011}, this galactic simulation includes a rotating frame of reference at a radius of 6 kpc. Gas at this radius does not move with respect to the grid, minimizing the artificial numerical support that occurs from circular motion over a Cartesian mesh. Gas to either side of the co-rotating radius moves in opposite directions. We therefore concentrate our analysis to within 1 kpc of the co-rotating radius, from $r > 5.5$\,kpc - $r < 6.5$\,kpc, and focus our refinement over this region, allowing it to incrementally decrease outside these radial points. This technique allows us to gain considerably more realistic spatial resolution for clouds that would otherwise become more bloated through such artificial numerical spreading.

\begin{figure}[!b]
\centering
\includegraphics[width=85mm,height=70mm]{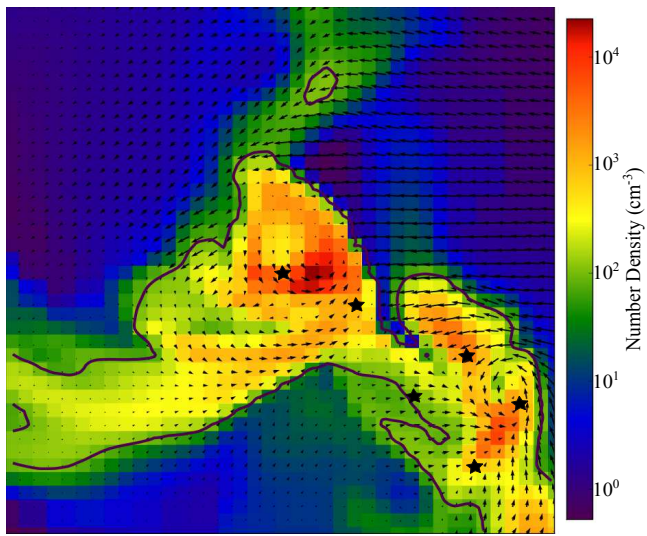}
\caption{A slice through the z-plane of a typical cloud.  The image size is $300$\,pc and overlaid with a contour to show regions within our number density threshold of $100$\,cm$^{-3}$, which would be tagged as belonging to the cloud.  The image is centred on cloud 1323, however, 5 other clouds also fall into the field of view (positions are marked with stars).  Velocity field vectors have also been overlaid to show the bulk motion of the gas. Note that the gas flows along the large-scale filamentary structure of the cloud and gathers in several dense regions (of order $10^4 cm^{-3}$).  Image rendered using yt \citep{yt}}
\label{fig:mod_velocity}
\end{figure}

Clouds in our simulations are identified using a friends-of-friends cell finding algorithm.  This process identifies clouds based on a density threshold, which we selected to be $n_{\rm thresh}$ = $100 {\rm cm}^{-3}$. This value was selected based on a review of
the observational catalogues of Milky Way GMC's \citep{rathborne_clump, duval}, as being a representative value for the average density of these objects.   

The algorithm initially identifies peak cells in the simulation which have a density higher than $n_{\rm thresh}$ and exceed the density values of their neighbouring cells. If two peak cells are closer than a set separation length, $l_s = 4\times \Delta x$, where $\Delta x$ is the minimum cell size of 7.8\,pc, then they are tagged together as the same cloud. Neighbouring cells whose density also exceeds $n_{\rm thresh}$ are then assigned to the cloud via a friends-of-friends scheme, whereby the code steps through the neighbours of each cell assigned to the cloud to search for more cloud components. A more detailed description of this process can be found in \cite{Tasker2009}. 

Clouds in this analysis are harvested from the torus enclosing the co-rotating radius, between $5.5$ to $6.5$\,kpc from the galactic center, where as discussed,  the artificial numerical effects will be minimized. We also limit our analysis to clouds that have at least three cells in each dimension; a minimum of 27 cells in total. This ensures our objects have our best resolution and are suitable candidates for the simulated catalogue. The full criteria for the catalogue are discussed in Section~\S6. 


\section{Structure of the Clouds}
\label{sec:structure}

Figure~\ref{fig:zoom_scale} shows density projections of the face-on disk at different scales. From left to right, the first panel shows a 20\,kpc spread of the entire disk. The co-rotation radius is clearly visible as the position of the most resolved clouds, which have access to all refinement levels and are unhindered by artificial support from motion over the grid. To aid the eye, white rings have been overlaid over our analysis region at disk radii of 5.5\,kpc and 6.5\,kpc.  The center panel zooms into a 5\,kpc portion of the region of co-rotation.  Here again it is clear that the most resolved clouds are found in the region of co-rotation.  The last 1 kpc panel zooms in on two clouds tagged by the cloud finder algorithm, which are centrally positioned in the image.  The differences between the structures well outside the high-resolution, co-rotating region are due to differences in resolution and numerical diffusion.  The co-rotation technique dramatically reduces numerical diffusion due to advection errors.  The combined effect is roughly equivalent to convolving the high resolution data with a 50 pc Gaussian filter.  As shown in Figure \ref{fig:zoom_scale}, the highly resolved chains of clouds connected by thin filaments between 5.5 and 6.5 kpc become single thick filaments at low resolution (near 7 kpc) and the cloud complexes become large clouds with the same total mass.

A $300$\,pc wide slice through a typical cloud identified by the algorithm is shown in Figure~\ref{fig:mod_velocity}.  The cloud is elongated, with clear filamentary structures created through tidal interactions with nearby clouds. It is also flattened, having most of its extent in the xy-plane.  The contour line shows cells that would be tagged by the algorithm at the density threshold of $100$\,cm$^{-3}$.  We can see from the overlaid velocity fields that most of the gas in the cloud and surrounding interstellar medium is flowing into the densest knot of the structure, within which would lie the gravitationally collapsing core.  The gas flow is strongly directed along the filaments, towards the dense regions within the GMC. 

\begin{figure}
\centering
\includegraphics[scale=0.35]{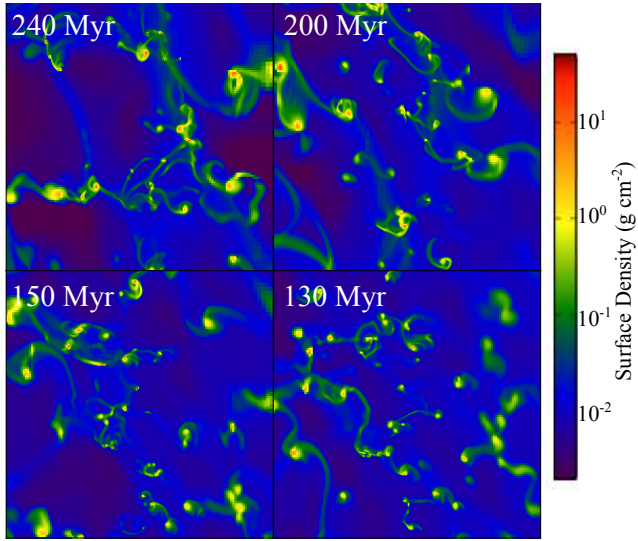}
\caption{Surface density maps of the same portion of the disk taken at 4 different time steps.  Each of the images span 2 kpc.}
\label{fig:evol}
\end{figure}

Shown in Figure~\ref{fig:evol} is a density projection of a 2 kpc portion of the galactic disk, shown at four different times in the simulation; 130, 150, 200 and 240\,Myr. Our analysis is completed for the clouds in the disk at 240 Myr, after one complete orbital period for the edge of our main region at $r = 6.5$\,kpc. These frames allow us to see the evolutionary paths that our clouds follow.  Early on at 130\,Myr, the structures formed are very small and almost bead-like.  As times passes, these smaller objects grow larger through processes of collision and gas accretion. Once more massive, the continuing interactions between the clouds lead to spiral tails of lower density gas being stripped from the outer layers. By 200 Myr the clouds have taken on the more expected filamentary nature shown in Figure~\ref{fig:mod_velocity}.

\section{Analysis of Global Cloud Properties}
\label{sec:bulk}

\begin{figure*}
\centering
\begin{tabular}{cc}
\includegraphics[width=70mm,height=55mm]{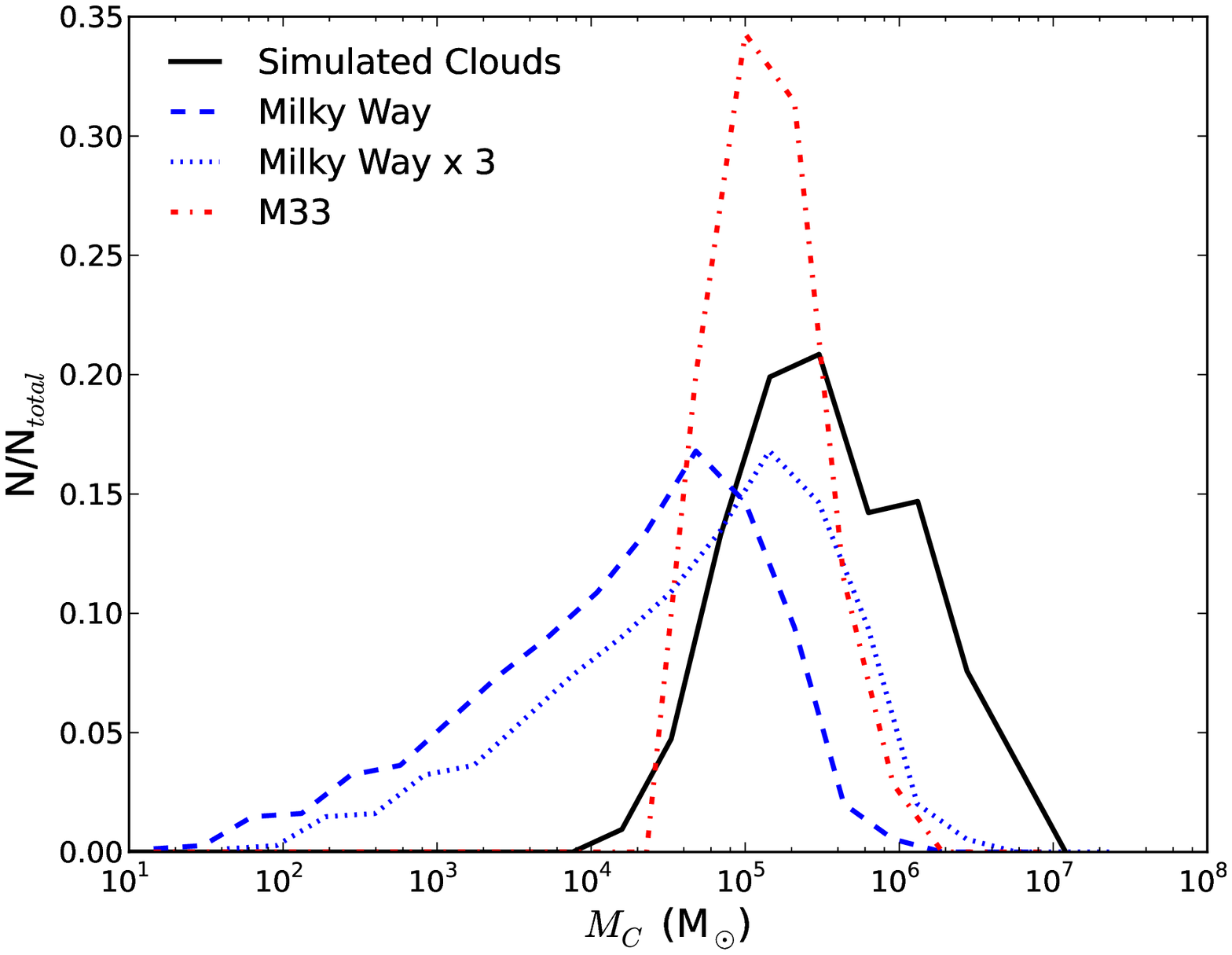}&
\includegraphics[width=70mm,height=55mm]{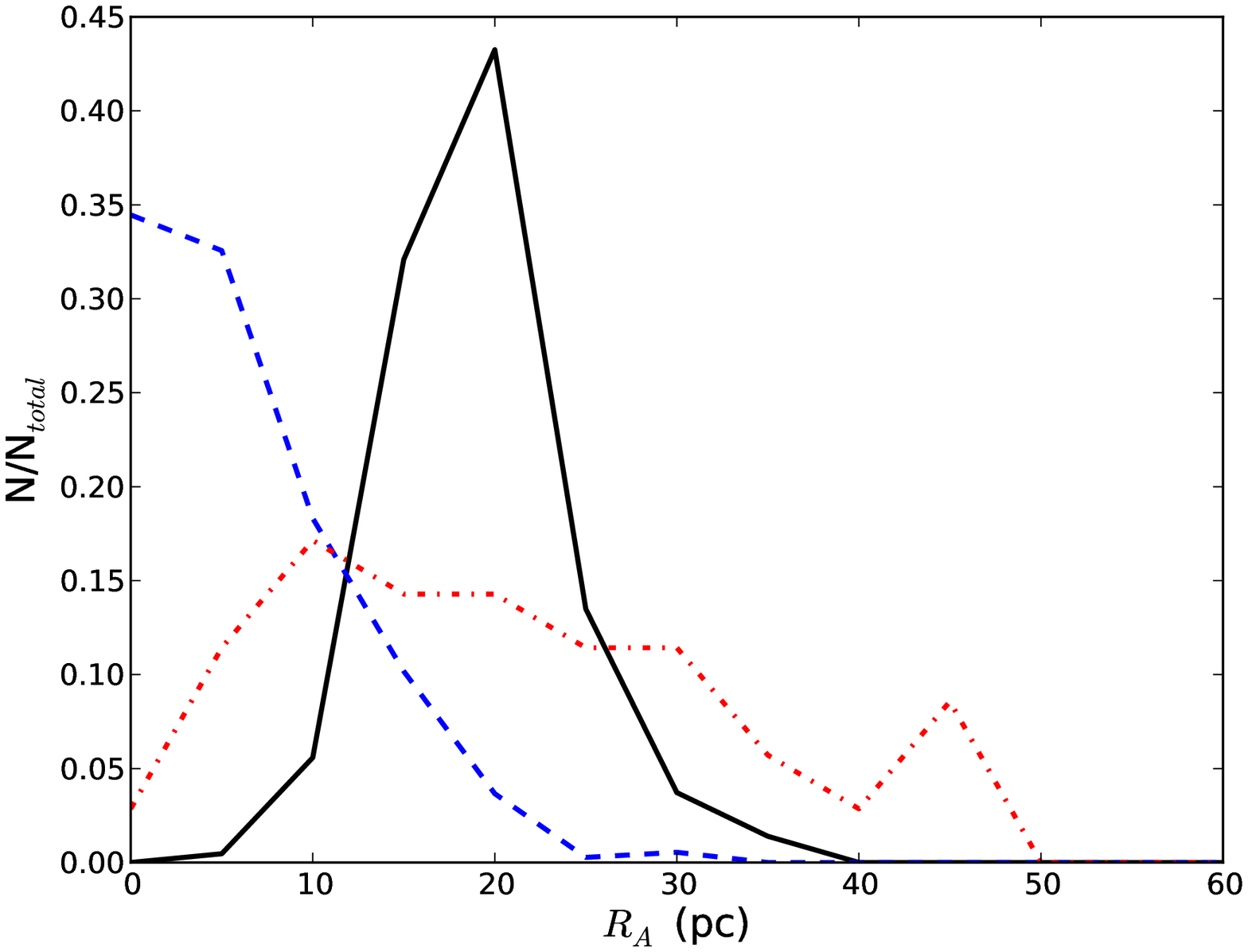}\\
\includegraphics[width=70mm,height=55mm]{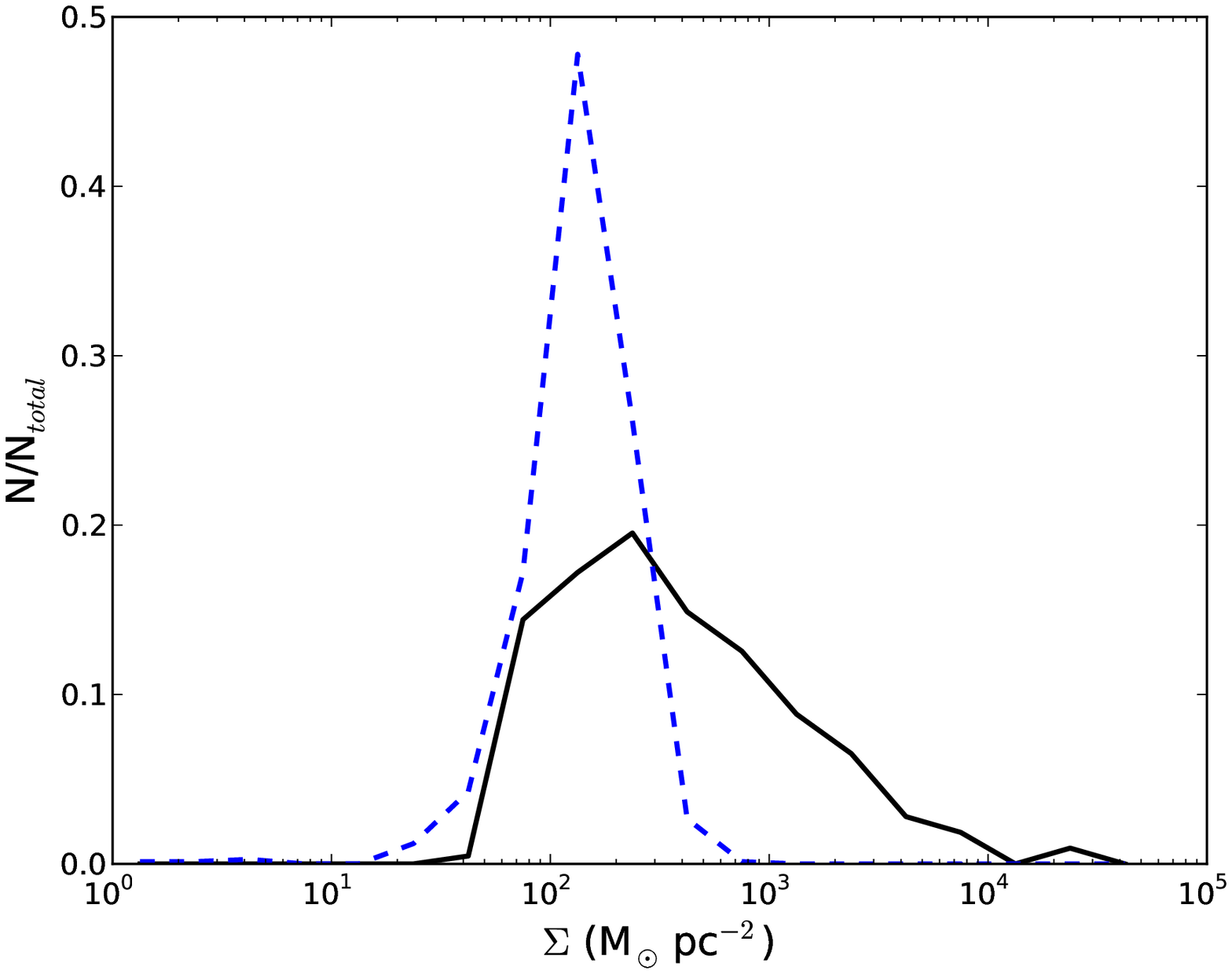}&
\includegraphics[width=70mm,height=55mm]{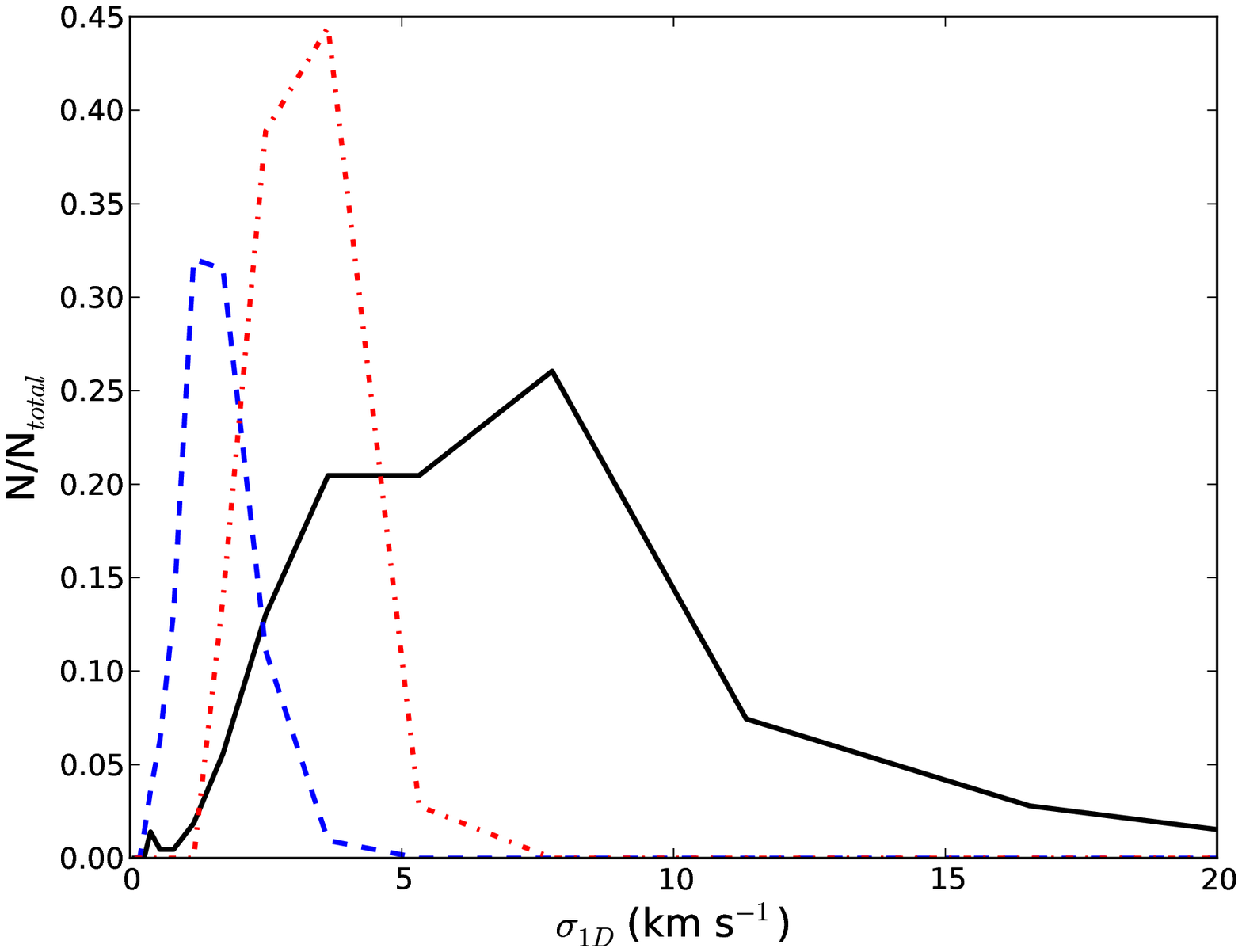}\\
\includegraphics[width=70mm,height=55mm]{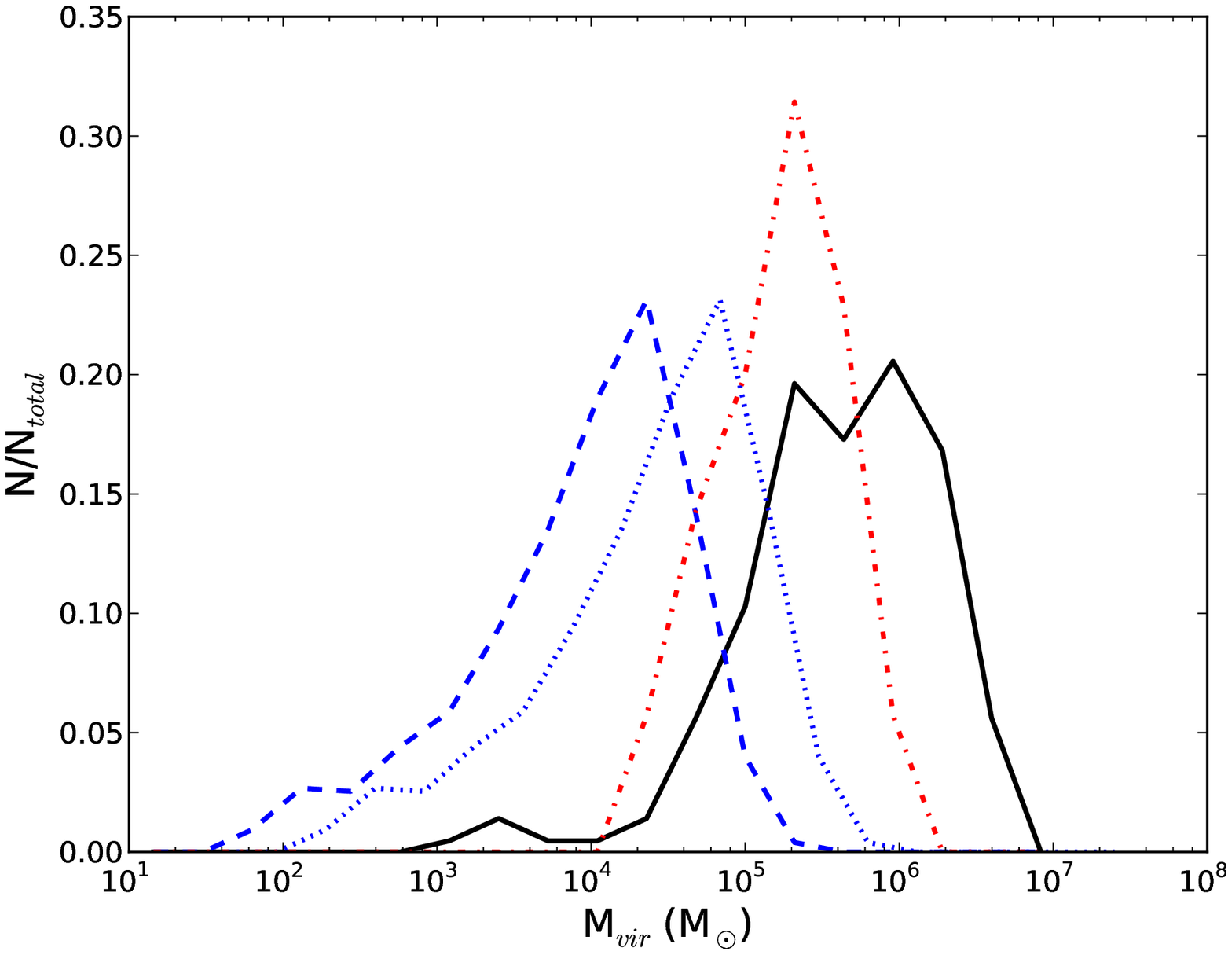}&
\includegraphics[width=70mm,height=55mm]{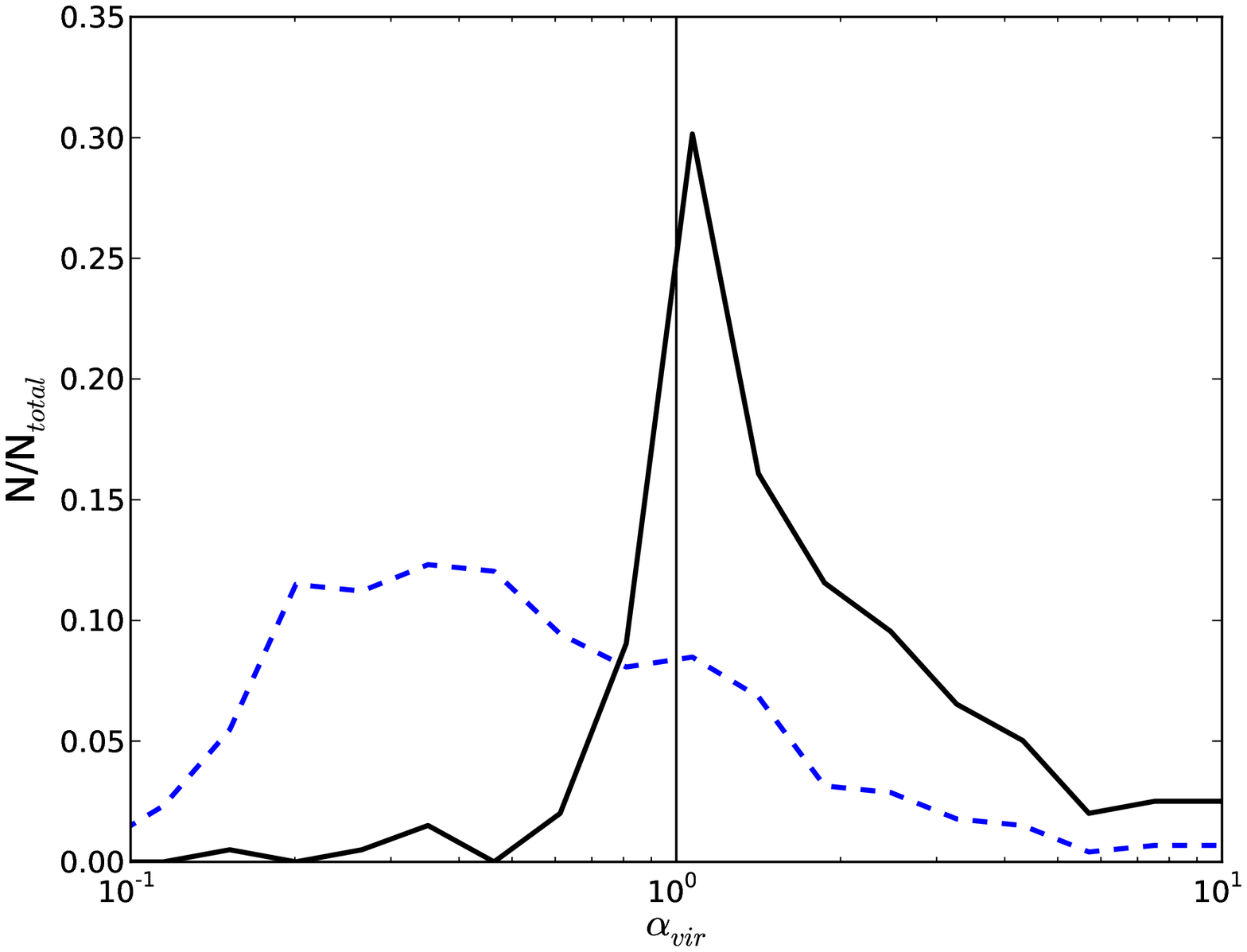}\\
\end{tabular}
\caption{The global cloud property distributions.  Top row shows cloud mass (left) and cloud radius calculated from the projected area, $R_A \equiv \sqrt{A_{yz}/\pi}$ (right ).  Middle row shows surface density based on the in-plane cloud area, $\Sigma \equiv M_c/A_{yz}$ (left) and the 1D velocity dispersion, $\sigma_{\rm 1D} = 1/\sqrt{3}\left(v_x^2+v_y^2+v_z^2\right)^{1/2}$ (right). Bottom row shows the virial mass, $M_{\rm vir} =  5(c_s^2+\sigma_{\rm 1D}^2)R_A/G$ (left) and virial parameter, $\alpha_{\rm vir} = 5(c_s^2+\sigma_{\rm 1D}^2)R_A/GM_c$ (right). Our simulated clouds (black solid line) have a minimum of 3 cells in each $x, y, z$ dimension and live in our maximally resolved torus around the rotating frame of reference, $5.5 < r < 6.5$\,kpc. These are compared to the clouds in the GRS Milky Way survey \citep{duval} (blue dashed line) and the M33 survey\citep{M33} (red dot-dashed line).  In each of the mass panels, the results for Milky Way masses multiplied by 3 (blue dotted line) are shown.  }
\label{fig:histograms}
\end{figure*}

Figure~\ref{fig:histograms} presents the bulk physical properties of the simulated GMCs at $t = 240$\,Myr. The six histograms show the results for the cloud mass, surface density, velocity dispersion, virial mass and virial parameter. The dashed overlaid lines show results from the observational data sets in the GRS Milky Way survey (blue dashed) and the M33 survey (red dot-dash). 


The top left histogram in Figure~\ref{fig:histograms} shows the cloud mass as calculated by the total mass in all cells associated with the cloud via our cloud finder scheme described in Section~\S2. This cloud mass is compared to the derived CO mass of the GRS Milky Way and M33 GMCs. Comparing our result with the two observational data sets, we see the peak in our distribution agrees well with the M33 peak mass at $M_{\rm c, peak} = 10^5$\,M$_\odot$ . With the M33 mass range extending from $3\times10^4 $\,M$_{\odot}$ to $1\times 10^6$\,M$_{\odot}$, we find good agreement with the overall profile, with our own mass range stretching between  $5.7\times10^4$\,M$_{\odot}$ to $2.7\times10^7$\,M$_{\odot}$ \citep{M33}. The Milky Way data set finds a slightly lower peak mass at $M_{\rm MW, peak} = 4.8\times10^{4}$\,M$_\odot$, reflecting the order of magnitude higher spatial resolution of the  GRS Milky Way survey. However, it is worth noting that the Milky Way data cannot be quoted without some error. \citet{Heyer2009} reports (private communication) that the masses of the Milky Way may be larger by a factor of 3 or more.  If this proves to be true, our mass range is in good agreement with both data sets. 

While the majority of our clouds seem to agree well with the observational results, we do find a high mass tail with clouds extending to masses higher than that found in either the Milky Way or M33.  This is due to the lack of star formation and feedback in the simulation, which would inhibit cloud growth \citep{Tasker2011}. Without such destructive mechanisms, clouds can continue to grow via accretion and mergers throughout the simulation, forming a population of old, tightly bound clouds that in reality would have formed a stellar cluster. We will discuss these objects in Section~\S 5 and, since we do not expect to find them in actual galaxies, they are not included in the cloud catalogue. 


The top right histogram shows the distribution of cloud radii. We define the radius of a cloud via:
\begin{equation}
R_A = \sqrt{\frac{A_{yz}}{\pi}}
\end{equation}
where $A_{yz}$ is the projected area of a cloud in the yz-plane of the disk, an orientation chosen to agree with the typical observed angle along the plane of the disk, as in the GRS Milky Way survey. A similar definition of the radius is used for both of the comparison observational surveys.  We have compared our findings using an area projected from the xy- and yz- planes and found no significant change to our cloud properties ($A_{yz}$ is used for the remainder of the analysis).  Again, we find our cloud radii correspond well with the M33 data set, although our most common radius falls slightly higher than the M33 clouds; at $R_A = 20$\,pc, compared with $R_{\rm M33} = 10$\,pc. The range of radii for our clouds falls between that of the Milky Way data and M33, with a spread of $40$\,pc. This is a smaller range than what is found for M33, possibly due to the more uniform nature of our cloud environment since we draw clouds from a $1$\,kpc wide torus in the center of our disk, which is unaffected by galactic structural differences. 


In the second row on the left, we show the mass surface density of the clouds, which is defined as $\Sigma \equiv M_c/A_{yz}$. For this property only Milky Way data is available, which peaks at a comparable value to our clouds at $\Sigma_{\rm peak} \sim 100-200$\,M$_\odot/{\rm pc}^2$. Notably, we get good agreement for the surface density without any form of feedback, suggesting that internally supported turbulence in clouds may be driven by cloud interactions, rather than localized energy from star formation. 


The right-hand histogram on the second row of Figure~\ref{fig:histograms} shows the clouds' internal velocity dispersion. The value used is the mass-weighted one-dimensional velocity dispersion about the cloud's center of mass velocity, calculated as:
\begin{equation}
\sigma_{1D} = \frac{1}{\sqrt{3}}\sqrt{v_x^2+v_y^2+v_z^2}.
\end{equation}
Here we obtain a wide range of velocity dispersion values, upto a maximum of $14$\,km/s with our most common value just below $8$\,km/s. This peak is slightly higher than the Milky Way and M33 clouds, which find peak distribution values of $1.1$\,km/s and $3.6$\,km/s respectively and is more in keeping with the velocity dispersion of the ISM \citep{Kennicutt1998}. 
The velocity dispersion of the clouds is likely to arise from two effects. The first of these is a dispersion driven by cloud-cloud interactions. As Figure 3 shows, the cloud environment is rife with structural interactions and this affects all the clouds, regardless of their size. The second cause is for the small tail population of massive clouds seen in Figure 4. These objects have a deeper potential well which allows an increased velocity dispersion, likely to be seen as the spread of higher values in our distribution. This may decrease with the introduction of stellar feedback, without which the clouds may become more gravitationally bound for a given size than their observed counterparts. We will consider this again in Section~\S 5.


Our bottom two plots show cloud properties related to their virialisation. A measure of this is the virial parameter, $\alpha_{\rm vir}$, which is related to the ratio of kinetic to gravitational energy of a cloud. \citet{Bertoldi1992} define this as:
\begin{equation}
\alpha_{\rm vir} = \frac{5(c_s^2+\sigma_{\rm 1D}^2)R_A}{GM_c}
\label{eq:alpha}
\end{equation}
where we include both the thermal ($c_s$) and non-thermal ($\sigma_{\rm 1D}$) contributions to the cloud's velocity dispersion. $R_A$ is the radius as calculated from the project area, as above, and $M_c$ is the cloud mass. 

For a uniform, spherical cloud, a value of $\alpha_{\rm vir} = 1$ implies that the kinetic energy of the cloud is half its gravitational potential, making clouds with $\alpha_{\rm vir} < 1$ dominated by gravity. If we assume that clouds are virialised objects with $\alpha_{\rm vir} = 1$, we can reverse Equation~\ref{eq:alpha} to get an expression for the {\it virialised mass}:
\begin{equation}
M_{\rm vir} =  \frac{5(c_s^2+\sigma_{\rm 1D}^2)R_A}{G}
\end{equation}
This value is plotted on the left-hand histogram on the bottom line of Figure~\ref{fig:histograms}. Our clouds agree well with the M33 data set, although we do see a population of high mass clouds as we did for the mass distribution, due to lack of star formation to inhibit the formation of large clouds. The higher resolution Milky Way data shows a smaller mass population which we might expect to match closer if we could resolve our clouds at the same spatial limit.

The right-hand plot on the bottom row shows the value of $\alpha_{\rm vir}$ for the clouds, plotted with the Milky Way data (there are no M33 values available for this quantity). Our clouds are less bound than those identified in the Milky Way: we find $15$\% of clouds have $\alpha_{\rm vir} < 1$, whereas the GRS Milky Way survey finds approximately $70$\%. This is in contrast to findings by \citet{Dobbs_virial} and \citet{hopkins} that a lack of feedback produces a more bound population.  One explanation might be that the higher resolution observed Milky Way data is identifying bound subregions within our clouds.  Alternatively, the less bound population may be destroyed by feedback effects.


\begin{figure}
\centering
\includegraphics[width=80mm,height=60mm]{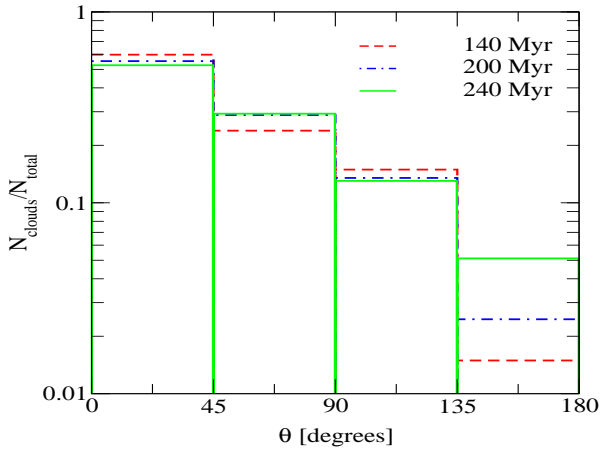}
\caption{Distribution of the angle of the cloud angular momentum vector with respect to that of the galaxy at three different simulation times, $t = 140$ (red dashed), $200$ (blue dot-dash) and $240$\,Myr (green solid). At early times, more clouds show a prograde rotation, with $\theta < 90^\circ$, rotating in the same sense as the galaxy. As the simulation progresses, a retrograde population develops.}
\label{fig:orientation}
\end{figure}

Figure~\ref{fig:orientation} plots the distribution of directions of  cloud angular momentum vectors with respect to the galaxy. While all clouds move predominantly through the galaxy with the rotation of the disk, their rotation around their own axis may differ. Clouds with a prograde rotation have $\theta <  90^{\circ}$ and rotate in the same sense as the galaxy while clouds with a retrograde rotation have $\theta >  90^{\circ}$ and rotate in the opposite direction. In M33, \citet{M33} found that clouds show a slight preference for aligning with the galactic rotation, with $60$\% of clouds showing a prograde rotation having $\theta < 90^\circ$.

At early times, the clouds in our disk are born prograde. This is shown in Figure~\ref{fig:orientation} by clouds at $t = 140$\,Myr having the highest fraction of clouds with $\theta < 45^\circ$. At later times, a retrograde population develops as new clouds feel not only the disk shear (which from angular momentum conservation must produce a prograde cloud) but also the gravitational pull of neighbouring clouds. The clouds also undergo collisions and interactions during their lifetime which may result in either a prograde or retrograde rotation. By $t = 240$\,Myr, $18$\% of the clouds rotate retrograde. This is a smaller fraction than those observed by \citet{M33}, but in agreement with previous simulations performed by \citet{Tasker2011}. In their paper, \citet{Tasker2011} find that the addition of photoelectric heating reduces the velocity dispersion in the disk ISM, resulting in a smaller retrograde population. This result was also noted by \citet{Dobbs_agg}. The addition of feedback is likely to affect this result, providing an additional force that can increase the disk velocity dispersion.  Regardless of the exact value of the retrograde percentage, the existence of such a population strongly suggests that the cloud's external environment plays a key role in its evolution, providing the necessary forces to create the spread of orientations observed in the GMC surveys. 

\subsection{Mass fraction of star forming gas in clouds}

\begin{figure}
\centering
\includegraphics[width=80mm]{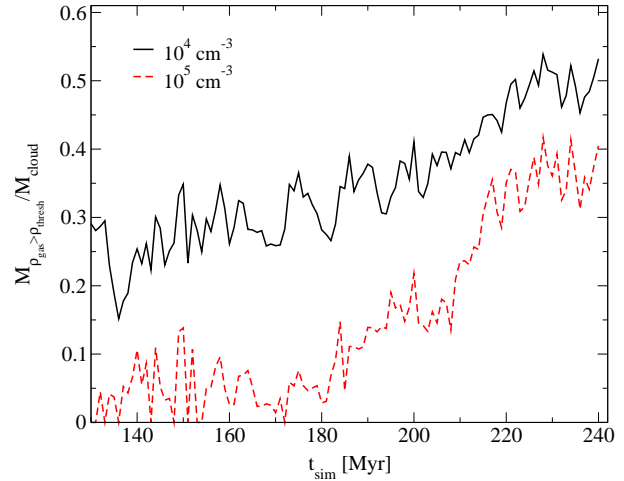}
\caption{Fraction of gas mass within the GMCs that is higher than the density thresholds, $n_{\rm thresh} = 10^4$\,cm$^{-3}$ and $n_{\rm thresh} = 10^5$\,cm$^{-3}$. $10^4$\,cm$^{-3}$ is the threshold at which star formation is observed to occur.}
\label{fig:massfrac}
\end{figure}

At $100$\,cm$^{-3}$, the density we use to define a GMC refers to the whole cold gas cloud, but does not give us a strong handle on how much star formation we should expect. Figure~\ref{fig:massfrac} shows the fraction of mass contained in clouds that is higher than two density thresholds, $n_{\rm thresh} = 10^4$\,cm$^{-3}$ and $n_{\rm thresh} = 10^5$\,cm$^{-3}$, a factor of 100 and 1000 times higher than the cloud definition threshold.  The value $ 10^4$\,cm$^{-3} $ is the density of clumps in which star formation is observed to occur \citep{Ginsburg2012, Lada2010}. As simulation time progresses, the density of gas within the clouds increases. This agrees with what we saw visually in Figure~\ref{fig:evol}; small objects at early times gain mass through accretion and mergers, which causes the clouds to get more dense.  We will also see this in the next section when we look at the evolution of the mass profile over time.  

This result provides interesting insights into how molecular clouds set themselves up for star formation.  
The Figure shows that between $t = 130 - 200$\,Myr, approximately 30\,\% of the cloud mass is at densities higher than $10^4$\,cm$^{-3}$. At this density, stars would be expected to form with an efficiency of around 10-30\,\% \citep{Lada2003}. This would give an overall cloud star formation efficiency of about 3-9\,\%. \citet{Krumholz2007} estimate that this number should be around $2$\%, implying that our clouds convert their gas to dense cores too efficiently. This is unsurprising, however, since there is nothing to prevent the collapse within the bound region of the cloud without the inclusion of local feedback. 

The rate at which star forming gas appears is also very interesting.  We can estimate from the approximately linear $n_{\rm thresh} = 10^4$\,cm$^{-3}$ curve, that the fraction of potentially star forming gas in a cloud grows by 30 \ \% over 100 Myr, or 3\ \% per 10 Myr, in clouds of roughly $10^6 M_{\odot} $.  This is in reasonable accord with star formation rates - suggesting that star formation rates in clouds are related to the rates at which gas can be accumulated into dense regions.  

A similar analysis can be performed that looks at surface, rather than volume, density thresholds. However, in this case, our result is dependent on our cloud selection scheme. Since clouds must be at least 3 cells in height, corresponding to a thickness of 23.4\,pc, and by definition must have a density $> 100$\,cm$^{-3}$, we have a sample-determined surface density threshold of $8\times 10^{21}$\,g/cm$^{-2}$, or a visual extinction of $A_v \simeq 10$. Since this is a result of our cloud sampling, the volume density threshold is the more revealing quantity. 

\subsection{Characterising cloud turbulence}
\label{sec:turb}

\begin{figure}
\centering
\includegraphics[width=80mm]{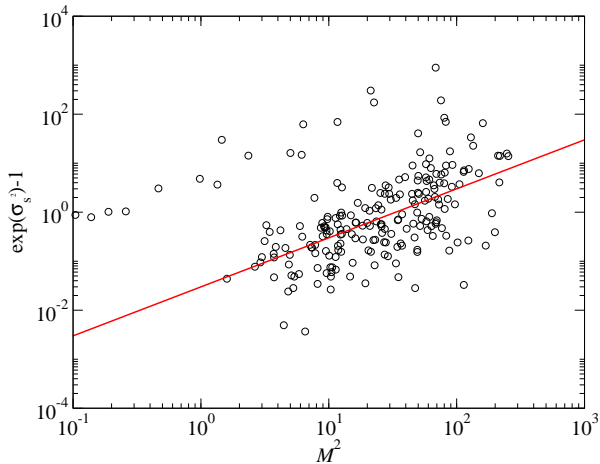}
\caption{Characterisation of the turbulence: red line fit corresponds to $y = 0.03 x$, giving a value for the turbulence parameter $b = 0.17$, suggesting that solenoidal turbulence is the dominant form in these simulations.}
\label{bturb}
\end{figure}

Another key to describing the internal structure of the GMCs is to investigate their turbulent energy by calculate their sonic Mach number, $\mathcal{M} \equiv <\sigma/c_s>$. For a non-magnetised gas, this value is linked to the variance in the logarithmic standard deviation, $\sigma_{\rm s} $ as:
\begin{equation}
\sigma_s^2 = \ln(1+b^2 \mathcal{M}^2)
\end{equation}
where $s = \ln(\rho/\rho_0)$ \citep{Kainulainen2013, Burkhart2012, Molina2012, Nordlund1999}. The numerical simulations in which turbulence is driven at large-scales have found that the value of the proportionality constant, $b$, depends on the mixture of solenoidal and compressive modes of the driving. In the case of fully solenoidal (divergence-free) driving, the value of the constant is $b=1/3$, and in the case of fully compressive driving it is $b=1$\citep{Federrath2010,Federrath2008}. Rearranging this equation gives:
\begin{equation}
\exp(\sigma_s^2)-1 = b^2 \mathcal{M}^2
\label{eq:bturb}
\end{equation}
which is plotted in Figure~\ref{bturb} with the left-hand side of Equation~\ref{eq:bturb} plotted on the $y$-axis and $\mathcal{M}^2$ on the $x$-axis. Since the plot is on logarithmic axes, we expect a gradient of $1$ and $(1,1)$ intercept at the value of $b$. The red line shows a $y = 0.03x$ fit, giving a value of $b = 0.17$, which is in agreement with the non-magnetized, isothermal simulations of turbulence that are driven with solenoidal forcing \citep{Federrath2010}. This suggests that the turbulence in the clouds of our simulations is not, on average, very compressive.

This finding agrees well with recent observational measurements of the value of $b$ for the infrared dark clouds in the Milky Way; dense molecular clouds that are thought to be likely candidates for future high mass star formation. This makes them particularly good comparison points for our non-star forming GMC population. For these objects, \citet{Kainulainen2013} find $b \sim 0.2$. Other observations of the Taurus and IC 5146 clouds suggest a slightly higher value at $b \sim 0.5$ \citep{Burkhart2012, Brunt2010, Padoan1997}.  In comparing simulations with cloud observations, \citet{Kainulainen2013b} suggest that non-magnetised clouds have $b < 0.3$ and clouds in the early stage of their evolution (the closest match with our population) are inferred to be lower still. While this agreement is pleasing, it is worth noting that our value for $\sigma_s$ is restricted by the densities that can be probed by the simulation. The true cloud may have a more extended range.

About a factor of 100 above our fitted line is a small population of clouds that appear to have a significantly higher value for $b$.  These clouds are old, tightly bound objects that greatly exceed their Jeans mass and sit in the high mass tail in Figure 8.  In reality, these objects would have already formed stars and so their presence here is a result of us not including star formation. As mentioned earlier in Section~\S 4, these objects are excluded from the cloud catalogue since we do not expect their structure to be found in actual GMCs.

\subsection{Evolution of cloud properties}

\begin{figure}
\centering
\includegraphics[width=75mm,height=60mm]{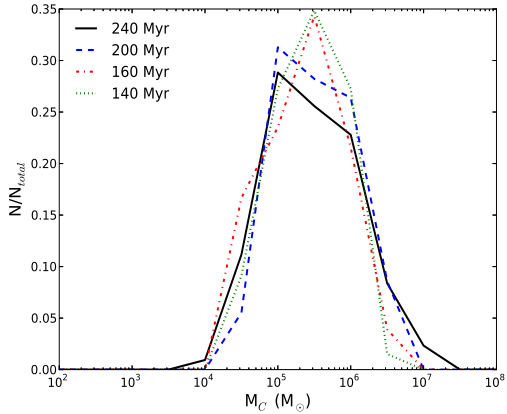}
\caption{Evolution of the cloud mass profile during the simulation. The profile is plotted at simulation times $140, 160, 200$ and $240$\,Myrs.}
\label{fig:mass_evol}
\end{figure}

One concern with measuring cloud properties in a numerical simulation is that they might depend on the simulation output examined. While we intentionally wait for a complete orbital period at our outer radius of interest before analyzing the cloud properties in Figure~\ref{fig:histograms}, it is also important to quantify any time evolution in our results. Figure~\ref{fig:mass_evol} shows the mass profile of the clouds at four different simulation times: 140, 160, 200 and 240\,Myr. 

The evolution of the cloud mass profile over the course of the simulation is small, with the range of cloud masses remaining approximately the same. Two small differences are seen at later times; the first and more important trend is the production of a high mass tail, with the maximum cloud mass extending from $4\times 10^6$\,M$_\odot$ at $t = 140$\,Myr to $2\times 10^7$\,M$_\odot$ by $t = 240$\,Myr. The second smaller trend is the shift in the mean cloud mass to lower values, changing from $6\times 10^5$\,M$_\odot$ at $t = 140$\,Myr to $2\times 10^5$\,M$_\odot$ by $t = 240$\,Myr. 

The major effect  --the high mass tail-- is due to the inability to destroy clouds via star formation or feedback. A small population of clouds continue to grow unabated via mergers and accretion to form a group of high mass, tightly bound clouds at the expense of lower mass clouds.  The second trend is likely due to the other end of the cloud's lifetime, where later simulation times see a population of clouds born along tidal tails and gas formed between cloud interactions. These clouds are smaller than those born out of the direct collapse of the disk.


\section{Scaling Relations}

Larson's empirical scaling laws relate the size, mass and velocity dispersion of molecular clouds \citep{larson1981}. The cloud radius and the velocity dispersion are found to obey a power-law relation of the form:
\begin{equation}
\sigma = \sigma_{\rm pc}R^{\alpha_o}
\end{equation}
where $\sigma_{\rm pc}$ and $\alpha_o$ are constants. While such a relationship is found to hold for both galactic and extra-galactic cloud populations, the exact value of these constants is debated. In his seminal paper, \citet{larson1981} finds values of $\sigma_{pc}=1.1$ and $\alpha_o=0.38$ for clouds in the Milky Way, which \citet{solomon} remeasures at $\sigma_{pc}=0.72\pm0.07$ and $\alpha_o=0.5\pm0.05$. GMCs in M33 yields values of $\sigma_{pc}=0.72$ and $\alpha_o=0.45\pm0.02$ \citep{M33, bolatto}. 

\begin{figure}
\centering
\includegraphics[width=90mm,height=75mm]{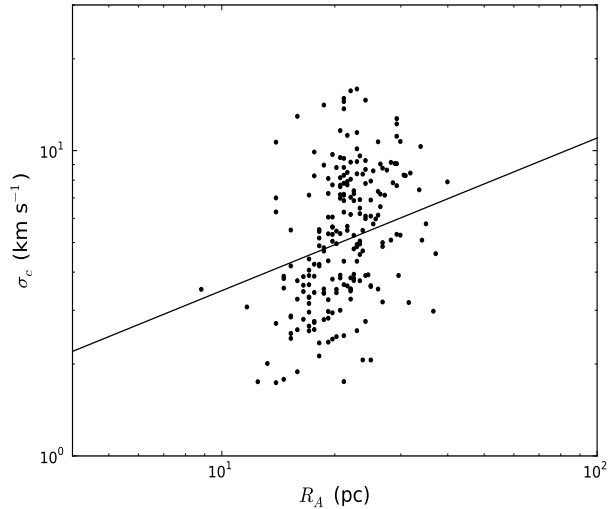}
\caption{Larson's size-linewidth relation, $\sigma = \sigma_{pc}R^{\alpha_o}$.  The black line shows Larson's original fit, with $\sigma = 1.1R^{0.38}$ and the individual points show our simulated clouds. }
\label{larson_rv}
\end{figure}

In Figure~\ref{larson_rv}, we show the line for the original Larson fit, $\sigma = 1.1R^{0.38}$, over-plotted with the values for each of our clouds in our main torus. On the y-axis, we use the combined thermal and non-thermal velocity dispersion components, $\sigma_c \equiv (c_s^2 + \sigma_{\rm 1D}^2)^{1/2}$. Our clouds lie on the fitted line, but show a large scatter over a small radius range. The small variation in radius is due to the region in which we harvest the clouds in the simulation; we intentionally take them from a uniform 1\,kpc thick torus in the disk and do not expect as wide a spread in cloud properties as if we had sampled through a larger region with a galactic density gradient. As discussed in Section~\ref{sec:bulk}, clouds in our simulation have a wider velocity dispersion than those observed in the Milky Way and M33, due to lack of feedback. This also impacts our scaling relations, giving a wide scatter of velocity dispersions for a given radius. 

\begin{figure}
\centering
\includegraphics[width=90mm,height=75mm]{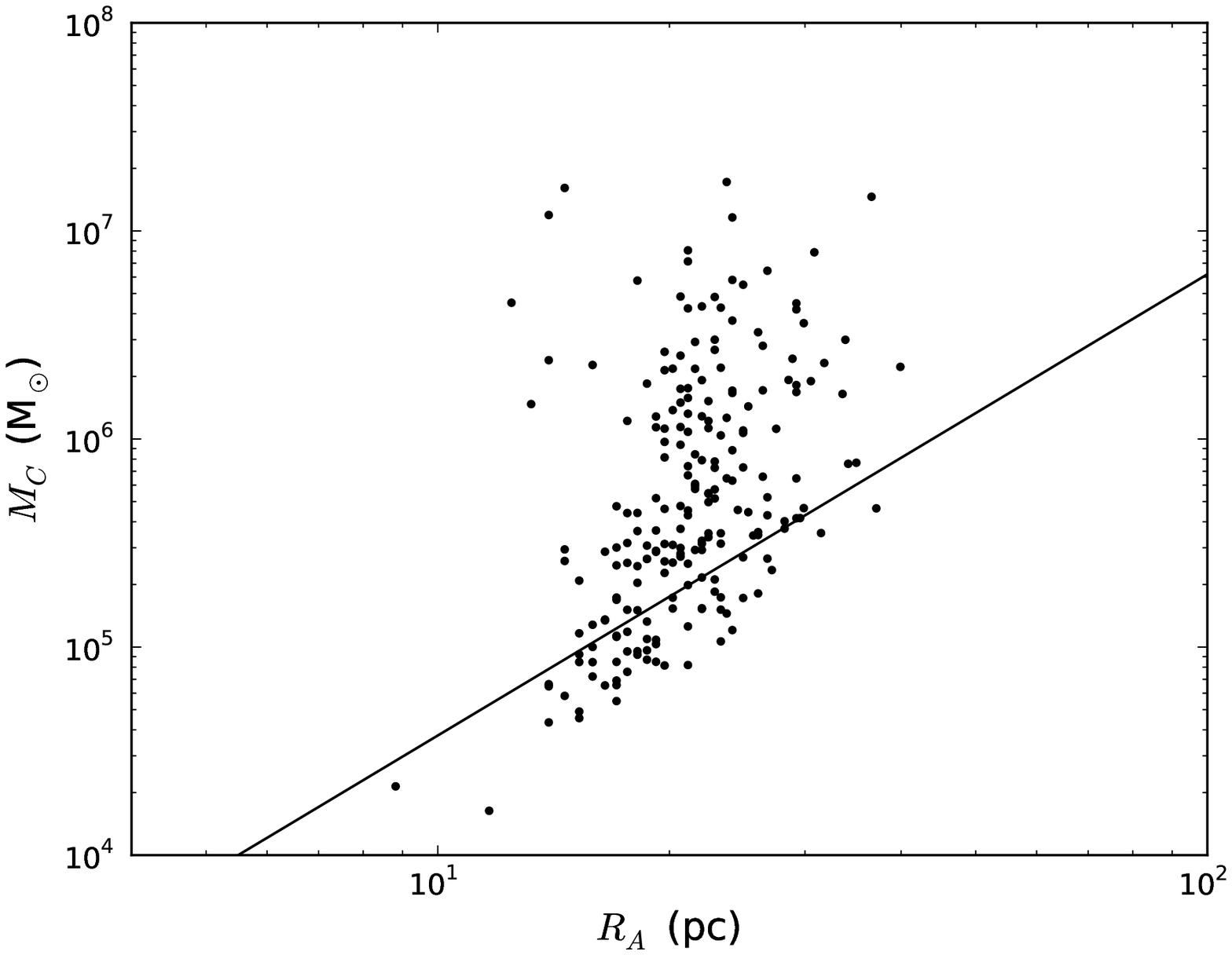}
\caption{Size-mass relation, $M = M_{pc}r^D$. The black line shows the fit to the Milky Way data in \citet{duval}, with $r = 228M^{2.36}$ and the individual points show our simulated clouds. }
\label{rm_fractal}
\end{figure}

A second scaling relation can be derived by considering the GMCs to be fractal structures. The fractal dimension, $D$, can be defined as the relation between the mass and radius of the studied bodies \citep{mandelbrot}:
\begin{equation}
M \propto R^D
\end{equation}
For clouds in the Milky Way, the fractal dimension had been measured as $D = 2.36\pm 0.04$ \citep{fractalpaper, duval}. The result for our simulated clouds is shown in Figure~\ref{rm_fractal},  which again shows a large scatter.  

\section{A Catalogue of Simulated GMC's}

\begin{deluxetable*}{lllllllllll}
\centering
\tablewidth{290pt}
\tablecaption{Catalogue of Simulated GMC Global Properties}
\tablehead{
\colhead{Tag}&\colhead{x}&\colhead{y}&\colhead{z}&\colhead{$R_{gal}$}&
\colhead{$\sigma_{1D}$}&\colhead{$R_A$}&\colhead{$M_c$}&\colhead{$\langle n\rangle$}&\colhead{$\Sigma$\tablenotemark{$\dagger$}}&\colhead{$\alpha_{vir}$}
\\
\colhead{} & \colhead{(kpc)} & \colhead{(kpc)} & \colhead{(kpc)} & \colhead{(kpc)} & \colhead{(km s$^{-1}$)} & \colhead{(pc)} & \colhead{(M$_{\odot}$)} & \colhead{(cm$^{-3}$)} & \colhead{(M$_{\odot}$ pc$^{-2}$)} & \colhead{}
}
\startdata
15323&15.24&21.92&15.96&6.0&6.85&20.7&1.07$\times 10^6$&1159.1&797.0&1.12\\
18355&11.90&10.98&16.02&6.5&8.92&22.0&2.38$\times 10^6$&1094.2&1560.8&0.89\\
6059&11.90&11.05&16.01&6.4&11.81&30.5&4.12$\times 10^6$&1028.7&1406.9&1.23\\
6505&13.45&10.36&16.00&6.2&6.59&23.3&1.04$\times 10^6$&635.1&606.6&1.21\\
651&13.28&10.32&16.01&6.3&11.84&26.4&3.22$\times 10^6$&1467.0&1464.1&1.37\\
15406&13.36&10.32&16.00&6.3&8.77&26.8&1.27$\times 10^6$&446.7&564.2&1.96\\
13683&13.39&10.26&16.01&6.3&6.84&19.7&1.36$\times 10^6$&885.5&1112.4&0.84\\
13448&15.83&10.34&15.99&5.7&8.07&24.1&1.67$\times 10^6$&805.2&913.6&1.14\\
14621&15.82&10.46&16.01&5.5&9.29&24.1&2.70$\times 10^6$&1169.0&1473.8&0.93\\
15424&16.53&9.92&16.02&6.1&11.27&22.5&3.05$\times 10^6$&1623.4&1922.9&1.11\\
1073&10.90&13.66&16.01&5.6&14.14&23.3&4.15$\times 10^6$&2097.2&2428.1&1.33\\
229&10.50&15.16&16.03&5.6&12.61&24.9&4.04$\times 10^6$&1711.4&2066.1&1.17\\
19181&10.55&15.15&16.03&5.5&8.52&26.8&2.51$\times 10^6$&709.4&1113.6&0.94\\
10870&10.42&13.02&16.00&6.3&6.23&19.2&1.16$\times 10^6$&1276.4&998.2&0.81\\
8188&20.20&12.14&16.02&5.7&8.58&26.4&2.36$\times 10^6$&778.9&1075.5&1.00\\
14865&22.12&14.38&16.02&6.3&6.21&25.3&1.30$\times 10^6$&511.1&644.3&0.94\\
1200&22.10&15.23&15.99&6.2&10.13&28.6&3.22$\times 10^6$&1147.3&1254.6&1.09\\
13523&21.56&14.93&16.02&5.7&5.45&18.2&1.06$\times 10^6$&1069.7&1020.0&0.65\\
270&21.60&15.00&16.03&5.7&7.08&22.9&1.36$\times 10^6$&1018.9&825.8&1.04\\
9668&20.75&12.61&16.02&5.8&13.00&23.3&4.27$\times 10^6$&2120.7&2496.2&1.09\\
12385&21.78&15.06&16.01&5.9&9.62&18.2&2.08$\times 10^6$&2210.9&2002.7&0.97\\
8396&21.94&15.00&16.01&6.0&9.20&20.2&2.12$\times 10^6$&1288.1&1657.1&0.97\\
1163&21.36&12.95&15.98&6.2&6.63&23.7&1.11$\times 10^6$&539.2&627.9&1.17\\
4640&21.30&12.88&15.98&6.2&6.83&21.6&1.00$\times 10^6$&677.1&685.9&1.24\\
8922&9.84&17.94&16.04&6.5&9.18&22.9&3.02$\times 10^6$&1274.0&1835.6&0.77\\
1323&9.85&17.36&16.02&6.3&11.43&29.2&1.75$\times 10^6$&664.8&653.1&2.59\\
12398&9.95&17.32&16.03&6.2&8.07&33.6&1.84$\times 10^6$&617.4&520.5&1.44\\
1838&9.98&17.29&16.01&6.2&11.15&28.9&5.40$\times 10^6$&2664.3&2058.9&0.79\\
10545&9.57&16.07&16.03&6.4&11.65&23.7&3.65$\times 10^6$&1878.0&2064.4&1.05\\
1475&21.16&18.28&16.01&5.6&12.18&29.6&4.12$\times 10^6$&1622.3&1499.9&1.26\\
17693&20.17&19.71&16.00&5.6&7.55&21.6&1.32$\times 10^6$&961.7&898.5&1.14\\
1452&22.36&16.97&16.01&6.4&8.49&27.2&1.70$\times 10^6$&819.5&734.1&1.39\\
11206&22.28&17.07&15.97&6.4&7.98&31.5&1.94$\times 10^6$&739.8&623.1&1.26\\
11692&22.00&17.61&16.02&6.2&7.12&21.6&1.45$\times 10^6$&762.8&987.4&0.93\\
6988&12.16&21.02&16.00&6.3&15.53&26.8&5.26$\times 10^6$&2057.5&2328.5&1.45\\
8434&16.09&21.73&16.02&5.7&8.51&26.8&1.69$\times 10^6$&788.9&748.8&1.39\\
19081&18.28&21.26&16.03&5.7&5.89&25.3&1.03$\times 10^6$&389.4&513.2&1.07\\
11191&18.31&21.30&16.03&5.8&11.54&23.7&2.70$\times 10^6$&1204.5&1526.0&1.39\\
16520&16.00&21.51&16.03&5.5&9.65&22.0&2.53$\times 10^6$&1534.0&1657.7&0.97\\
3358&18.93&21.67&15.99&6.4&7.86&34.1&1.93$\times 10^6$&761.7&528.1&1.33\\
8899&19.14&10.83&15.93&6.0&9.07&25.7&1.66$\times 10^6$&1228.8&800.6&1.53\\
17144&18.93&10.54&16.05&6.2&6.97&23.3&1.43$\times 10^6$&906.5&839.7&0.97\\
14354&17.64&10.13&16.06&6.1&8.96&22.5&1.76$\times 10^6$&1316.3&1107.9&1.24\\
13083&11.46&12.26&16.06&5.9&10.29&33.6&2.98$\times 10^6$&841.9&843.1&1.42\\
16541&21.67&13.86&16.04&6.1&8.23&22.5&1.97$\times 10^6$&1354.3&1238.3&0.94\\
1149&21.68&13.81&16.05&6.1&14.64&26.1&4.42$\times 10^6$&1676.3&2070.2&1.49\\
15150&11.69&20.19&16.02&6.0&7.90&30.9&1.99$\times 10^6$&619.1&665.6&1.18\\
14746&14.84&21.41&16.03&5.5&8.44&31.8&2.22$\times 10^6$&969.2&699.9&1.23\\
16961&17.00&21.58&16.00&5.7&6.56&36.3&1.72$\times 10^6$&538.1&414.8&1.13\\
1777&17.30&21.92&16.00&6.1&13.58&22.5&3.48$\times 10^6$&1989.1&2190.8&1.41\\
16128&17.21&21.86&16.07&6.0&8.31&18.7&1.32$\times 10^6$&1140.3&1198.0&1.19\\
\enddata
\tablenotetext{$\dagger$}{To quickly convert units: $\Sigma (g\ cm^{-2}) = 2.08\times 10^{-4}\Sigma (M_{\odot}\ pc^{-2})$ }
\tablecomments{Clouds in the catalogue lie within our most resolved torus, with galactic radii $5.5 < r < 6.5$\,kpc. They have a minimum of 3 cells in each $x, y, z$ direction and mass greater than $10^6$\,M$_\odot$.}
\end{deluxetable*}

A major goal of this project has been to systematically compare our simulated giant molecular clouds with observations and compile of a catalogue of the simulated clouds for further comparison and simulation. The clouds in the catalogue are presented in Table 1. This compilation includes each cloud's physical properties of mass ($M_c$), radius ($R_A$), average number density ($\langle n\rangle$), surface density ($\Sigma$), velocity dispersion ($\sigma_{1D}$) and virial parameter value ($\alpha_{\rm vir}$), along with their physical position in the disk. 

Clouds that are listed in the catalogue are considered our `best' clouds in terms of their numerical resolution. To belong in this category, clouds must fulfil the following conditions: (a) be found in our torus of best resolution, within the galactic radii $5.5 < r < 6.5$\,kpc. (b) consist of at least 27 cells, with 3 cells in each $x, y, z$ dimension. (c) have a mass greater than $M_c > 10^6$\,M$_\odot$. (d) Not fall in the over-dense, highly bound cloud population shown in Figure~\ref{bturb}. This corresponds to existing below the line $\exp(\sigma_s^2)-1 = 2\mathcal{M}$. These criteria ensure the cloud has the best internal properties our simulation can produce. When these criteria are applied, our highest mass cloud has $M_c = 5.4\times 10^6$\,M$_\odot$.

\begin{figure*}[]
\centering
\includegraphics[scale=0.35]{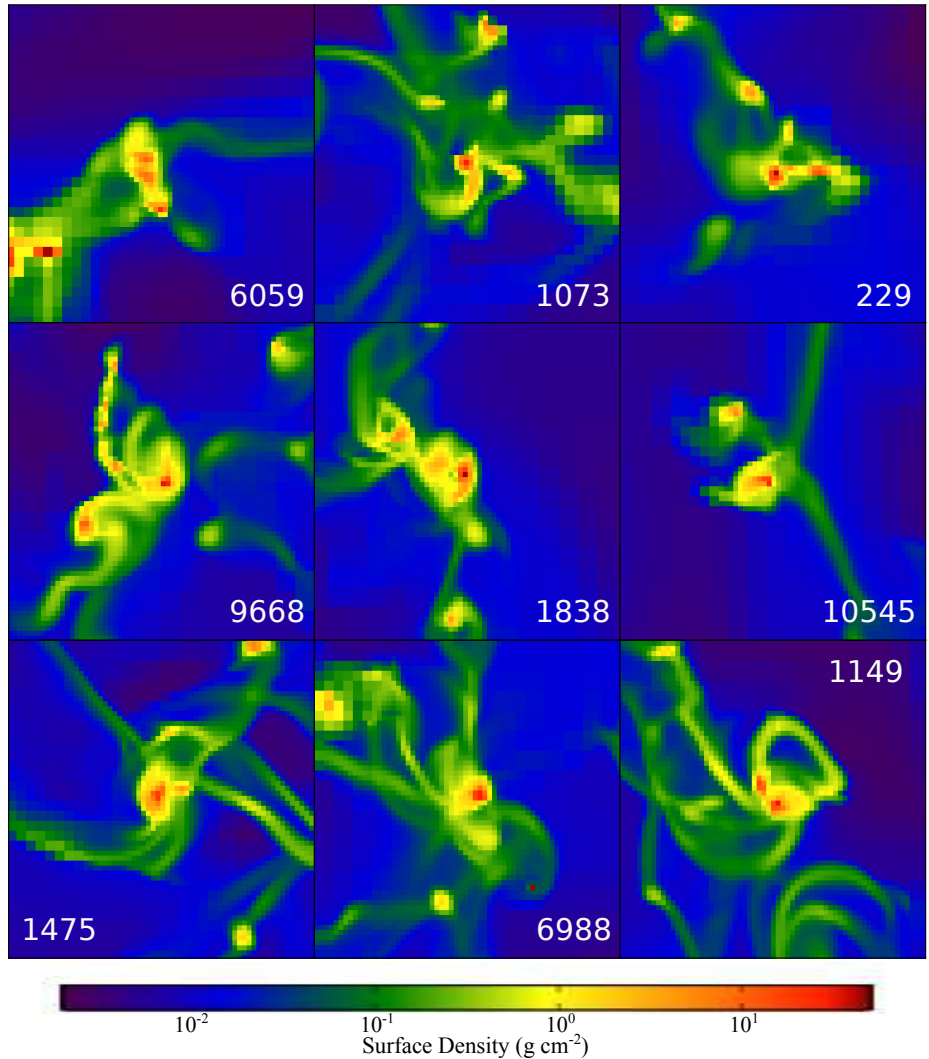}
\caption{Surface density maps of the 9 most massive clouds in the catalogue. Each image is 0.5 kpc across and is centred on the cloud center of mass.  Full set of catalogue images can be found on the accompanying website: http://www.physics.mcmaster.ca/mcclouds.
\label{fig:clouds1}}
\end{figure*}

Our simulated clouds are available from this site (www.physics.mcmaster.ca/mcclouds) in hdf5 format. The data consists of a three dimensional uniform grid $500$\,pc across, with each cell $7.8$\,pc on the side. The projected density of the 9 most massive of these cubes is shown in Figures~\ref{fig:clouds1}. As we discussed in Section~\S 3, visually our simulated clouds look similar to those observed \citep[e.g.][]{rathborne_clump, M33} with the elongated clouds shapes and a filamentary structure.  This suggests that the formation and evolution mechanisms at
work may be sufficient to produce the filamentary structure in the actual
galaxy.  Since we intentionally did not include an active star formation and feedback model in our simulation, the catalogue clouds should provide a good study for the earliest stages of the star formation process.

\section{Discussion and Conclusions}

We performed a high resolution simulation of an isolated galactic disk, which incorporated a rotating frame of reference at a galactic radius of $6$\,kpc. The use of the co-rotating frame improves the effective resolution by reducing the numerical dissipation.  GMCs were formed via a top-down gravitational fragmentation and were analysed between galactic radii $5.5 < r < 6.5$\,kpc, where their limiting resolution was $7.8$\,pc. In addition to the study of their properties, we present a catalogue of our best simulated clouds for further study of the formation of star clusters and high mass star formation. 

We find that our simulated clouds compare well with observed clouds both in the Milky Way and M33, even with the absence of feedback. Our typical cloud mass is between $10^5 - 10^6$\,M$_\odot$, in good agreement with the M33 data which is of comparable resolution to our simulation. We find a high-mass tail of clouds with masses up to $10^7$\,M$_\odot$ which builds up during the simulation due to the lack of stellar feedback. This lack of feedback also causes a larger range in velocity dispersion than is observed in either galactic survey. The radius, surface density and virial mass of our clouds all agree with the observational data sets we compared against, with our clouds being marginally unbound with $\alpha_{\rm vir} \sim1$.   The internal turbulence of the simulated clouds was estimated and found to be in agreement with solenoidal turbulence, in agreement with observations of clouds in the Milky Way. 

The fact that our simulated clouds are similar to observed ones is interesting because the cloud growth does not reach an equilibrium; clouds continue to grow with time.  The cloud growth phase, we argue, does mimic reality.  Molecular clouds do not live in isolation in the ISM but ought to be accreting mass from their surroundings.  They should also undergo a reasonable number of collisions before
their dispersal.  Clouds should therefore exhibit a significant range of gravitational boundedness
which is indeed observed in our simulations and in reality.  Even the neglect of star formation
feedback does not grossly affect cloud properties which suggests that cloud disruption need not
be catastrophic.

Our star formation efficiency, as estimated by the fraction of mass in clouds above a density of $n > 10^4$\,cm$^{-3}$, is between 3-9\%, a value higher than the observationally suggested fraction, but in keeping with our lack of feedback preventing any halt to the collapse of the bound core in the cloud.  
We also measured the rate at which dense, potentially star forming gas accumulates within dense
regions - wherein $n_{\rm thresh} \ge 10^4$\,cm$^{-3}$ -  to be  3\ \% per 10 Myr, in clouds of roughly $10^6 M_{\odot} $.  This again is in accord with 
observations of star formation rates in GMCs, and suggests that star formation rates are related to the rates at which dense gas can be accumulated into 
dense clumps in GMCs.  

The fact we were able to match many of the observed properties of the GMCs without the inclusion of star formation or feedback is proof of the importance of the galactic environment in which the GMC is born. Our results indicate that the galactic disk, with its rotational shear and interactions between neighbouring GMCs, play a vital role in the creation of the star formation environment. 

With this in mind, we present our cloud catalogue; a set of 51 simulated GMCs that can be used to explore the formation of stars in an environment that includes the full impact of the galaxy disk. We intend for these to be freely used by the star formation community to study this complex and fascinating topic. 



\acknowledgments
The authors would like to thank Helen Kirk, Mark Heyer and Jouni Kainulainen for many helpful discussions and the yt development team (yt: http://yt-project.org/) for their support during the analysis of this run.  We also thank the anonymous referee for comments that helped to improve the clarity of the paper. Numerical computations were carried out on Cray XT4 at Center for Computational Astrophysics, CfCA, of the National Astronomical Observatory of Japan. Analysis was performed on machines belonging to the Shared Hierarchical Academic Research Computing Network (SHARCNET:www.sharcnet.ca) and Compute/Calcul Canada. S.M.B.  was supported by McMaster University and E.J.T. would like to thank the Canadian Insitute for Theoretical Astrophysics who supported her as a CITA National Fellow. 
R.E.P.  and J.W. were supported by NSERC Discovery Grants. 
 

\end{document}